\documentstyle[preprint,tighten,aps]{revtex}

\newcommand{\resetcounter}{\setcounter{equation}{0}}     
\newcommand{\lagrange}{{\cal L}}

\newcommand{\A}  {\alpha}
\newcommand{\B}  {\beta}

\begin{document}

\draft
\preprint{UPR-734-T, hep-th/9703134}
\date{March 1997}
\title{Duality Invariant  Non-Extreme Black Holes in Toroidally Compactified
          String Theory}
\author{Mirjam Cveti\v c and Ingo Gaida
}

\address{Department of Physics and Astronomy \\
          University of Pennsylvania \\ 
      Philadelphia, PA 19104-6396, U.S.A.}
\maketitle\

\begin{abstract}
We present duality invariant structure of the thermodynamic quantities
of {\it non-extreme}  black hole solutions
of torodially compactified Type II ($M$-theory) and heterotic string   
in five and  four dimensions.  These quantities are parameterized by duality
invariant combinations of  charges and the non-extremality parameter, which
measures a deviation from the  BPS-saturated limit.
In particular, in $D=5$ we find explicit
S- and T-duality  [U-duality] invariant  expressions 
for solutions of  toroidally compactified heterotic string [Type II string].
In $D=4$, we  consider general S-duality invariant expressions for 
non-extreme solutions of 
pure $N=4$ supergravity  and find to the leading order in non-extremality
parameter the  T- and S-duality invariant expressions of
toroidally compactified  heterotic string.
General non-extreme solutions
 of toroidally compactified string in $D=4$ are awaiting further
investigation.
\end{abstract}



%
%

\pacs{04.50.+h,04.20.Jb,04.70.Bw,11.25.Mj}

\section{Introduction}

There have been a number interesting developments in understanding 
the black hole physics
in string theory  (for a review, see {\it e.g}, 
[\ref{horowitz},\ref{maldacena},\ref{cveticR}]). 
In particular,  string theory makes
it possible to address   microscopic  properties of black holes, like
the statistical origin of black hole
entropy and their radiation rates.  

The  prerequisite for such investigations  are the properties
of the classical black hole solutions.
A program to obtain the explicit form of the ``generating solutions''
for general rotating black hole solutions for toroidally compactified
heterotic and Type II string theories in diverse dimensions has been explored 
in detail (for a  review see  [\ref{cveticR}]). 
(Note that this study excludes notable  examples of  black hole solutions 
for $N=2$, $D=4,5$ string vacua; see {\it e.g.}, 
Ref. [\ref{luest}] and references therein.). Such string vacua possess enough (super)symmetries and  well understood 
moduli spaces, thus allowing for a reliable treatment 
of at least BPS-saturated solutions.

While the generating solutions specify the $D$-dimensional space-time, the
ultimate goal is to cast such solutions into {\it manifestly  duality
invariant} form.  This goal has been accomplished for  
BPS-saturated rotating black hole
solutions  in  $D=4$ and $D=5$  (in $D\ge 6$ these solutions have
zero area of the horizon), i.e. the 
Bekenstein-Hawking entropy and the ADM mass of 
BPS-saturated solutions of toroidally compactified  heterotic  
[Type II] string has been cast
in  S- and T-duality  [U-duality] invariant form.  
On the other hand, for the non-extreme solutions, primarily the structure of
the generating solution was given,  and  the duality invariant 
structure of {\it non-extreme}
solutions awaits further investigation (These issues have also been addressed
in Ref.[\ref{ortin},\ref{rajaraman}].).

The explicit form of the generating solutions
in $D=4,5$ of torodially compactified string was given 
in [\ref{cvetic1}-\ref{cvetic4}] (Particular examples of solutions have been 
obtained in a number of papers; for a review and references see 
[\ref{horowitz}] and references therein.).
The main purpose of this paper is to fill  in 
a gap by obtaining  a duality invariant form of 
 relevant thermodynamic quantities for  {\it non-extreme} solutions. 
These quantities are expressed in terms of duality invariant combinations of
(``dressed'') charges and the non-extremality parameter, which parameterizes a
deviation from the  BPS-saturated limit (This parameter can be in turn
traded for the ADM mass and a duality invariant combinations of  dressed
charges.).

In this attempt we were  successful for $D=5$  non-extreme solutions 
(for $D>5$ the procedure is
straightforward). In $D=4$  we 
were able to  obtain the results 
for certain truncations (``with  fixed moduli'') of $N=4$, $D=4$ supergravity
theory as well as  duality invariant expression to
the leading order in  the non-extremality
parameter for
toroidally compactified  heterotic string. Thus, 
the general non-extreme solutions
 of toroidally compactified string in $D=4$  still await further
 investigation.
The  stucture of these duality invariant  solutions may  also provide  
an indication about origin of  their microscopic  structure.

The structure of the paper is the following. In Section II.A  we start with 
the generating solution for the toroidally compactified heterotic (and
Type II string) in  $D=5$. In Sections II.B   and  II.C  we obtain 
general non-extreme duality invariant thermodynamic quantities for  $D=5$ 
black hole solutions in  toroidally compactified heterotic and
type II string, respectively. 
In  Section III.A we start with the four-charge parameter generating solution
of toroidally compactified heterotic string in  $D=4$. In  Section 
III.B we present non-extreme duality invariant thermodynamic quantities 
for general rotating black holes 
in pure  $N=4$ supergravity (i.e.  with ``fixed moduli'').
In Section III.C we discuss their
properties in  the extremal and BPS-saturated limits.
Finally, in Section III.D  we also derive  to the leading order in the
non-extremality paramter the  duality invariant 
expressions for  $D=4$ rotating black holes of toroidally 
compactified  heterotic string theory.
For the sake of completenes , S and T-duality transfomations of  torodially
compactified string in  $D=4,5$  are given in the  Appendix.

\section{N = 4 and N = 8 Black Holes in Five Dimensions}
\resetcounter

The  aim is to address general axi-symmetric solutions of  the bosonic 
sector of the effective Lagrangian of toroidally compactified string theory.
Since the bosonic sector includes the graviton, $U(1)$ gauge fields, 
as well as massless scalar fields,  
the axi-symmetric solutions correspond to the 
{\it dilatonic charged rotating black hole solutions}. 
According to the ``no-hair theorem'' these  black holes  in $D$-dimensions
are specified by the ADM  mass $M$,  
$[{{D-1}\over 2}]$-components of angular
momenta $J_{1,\dots, [{{D-1}\over2}]}$ 
and the number of allowed charge parameters associated with $U(1)$ 
gauge symmetry factors.

The most general black hole,
compatible with the no-hair theorem,
is  obtained by acting on the generating solution with 
classical duality transformations. 
 They do not change
the $D$-dimensional Einstein-frame metric but do change the charges and scalar
fields.  One  first  considers transformations,  belonging
to the maximal compact subgroup  of duality 
transformations on the generating solution, which 
preserve the canonical asymptotic values of the scalar fields 
and show that all charges are generated in this way.
Another  duality transformation is used to change the canonical
asymptotic values of the scalar fields to their arbitrary one. 
The aim is then to cast such a general solution 
in the manifestly duality invariant form.

We apply this method first to  $D=5$ {\it non-extreme}  
black holes of toroidally compactified string.

\subsection{The Generating Solution}

It has been shown in [\ref{hull}] that the generating solution 
for {\it both} the
toroidally compactified  Type II string {\it and} the  heterotic string 
can be specified by the $U(1)$ charges in  the NS-NS
sector. Specifically in $D=5$ the charge parameters of the generating
solution are: $Q_1^{(1)}, \ Q_1^{(2)}$ and  $P$. Here 
$Q_1^{(1)}, \ Q_1^{(2)}$ are the electric charges 
associated with  $U(1)$ gauge fields 
$A_{\mu\,i}^{(1,2)}$ of the (momentum, winding) sector of NS-NS sector of the
string theory, 
and $P$ is the  {\em electric} charge of the gauge field, whose field
strength is on-shell related to the field strength of the two-form field 
$B_{\mu\nu}$ by a duality transformation 
\footnote{In the following five dimensional case we take
$G_{N}= \pi/8$ and $g^{2} = e^{2\Phi_{\infty}/3}$.}. 
In addition the generating solution is specified by two angular momenta 
$J_{\psi,\phi}$ and the non-extremality parameter $r_0$, 
which measures a deviation from the BPS-saturated limit.  
The explicit form of this generating solution 
has been given in [\ref{cvetic4}].
The corresponding entropy reads
\begin{eqnarray}
 S &=& 4 \ \pi^{2}
\left \{ 
  \left [
          r_{0}^{3} \ ( 
\prod_{i=1}^{3}  \mbox{cosh}\delta_{i} + 
\prod_{i=1}^{3}  \mbox{sinh}\delta_{i} )^{2}
- \frac{1}{4} (J_{\phi}-J_{\psi})^{2}
  \right ]^{1/2}
+
\right.
\nonumber\\ & &
\left. 
\left [
          r_{0}^{3} \ ( 
\prod_{i=1}^{3}  \mbox{cosh}\delta_{i} - 
\prod_{i=1}^{3}  \mbox{sinh}\delta_{i} )^{2}
- \frac{1}{4} (J_{\phi}+J_{\psi})^{2}
  \right ]^{1/2}
\right \},
\end{eqnarray}
where the physical charges are
\begin{eqnarray}
\label{charge_def_5}
  P  &=& 2 \ r_{0} \ \mbox{sinh}\delta_{p} \ \mbox{cosh}\delta_{p}\ ,
\nonumber\\
 Q_{1}^{(1,2)}  &=& Q_{1,2} \ = \ 2 \ r_{0} \ \mbox{sinh}\delta_{q1,2} \
                            \mbox{cosh}\delta_{q1,2}.
\end{eqnarray} 
and the angular momenta are defined as:
\begin{eqnarray}
\label{non_mass_general_5}
 J_{\phi,\psi}  &=& 2 \ r_{0} 
\left ( 
 a_{1,2}  \prod_{i=1}^{3}  \mbox{cosh}\delta_{i} -
 a_{2,1}  \prod_{i=1}^{3}  \mbox{sinh}\delta_{i}
\right ).
\end{eqnarray} 
The ADM mass reads
\begin{eqnarray}
M  &=& r_{0} \ \sum_{i=1}^{3} (\cosh^2\delta_i+\sinh^2\delta_i) .
\end{eqnarray}
Here the non-extremality parameter $r_{0}$ and $a_{1,2}$
 parameterize respectively the ADM mass and the two components of 
the angular momentum 
of the   $D=5$  Kerr  (neutral, rotating)  solution  
(For  $r_{0}/2 \geq (|a_{1}|+|a_{2}|)^{2}$ 
the solution  has inner and outer horizon.).
The three charges are determined in terms of  $r_0$ and  
the  three ``boost'' parameters $\delta_i$, corresponding
 to the symmetry transformations  of stationary solutions 
 which generate charged  solutions from neutral ones  (for more details see
 [\ref{cvetic4}]).  

The BPS-saturated limit is obtained, by taking the non-extremality parameter
$r_0\to 0$ and the boost parameters $\delta\to \infty$, while keeping the
values of charges finite.

\subsection{Duality Invariant Black Holes of Toroidally Compactified Heterotic
String}

We start from the generating solution
of [\ref{cvetic4}] and a scalar non-linear $\sigma$-modell
with coset space 
$SO(1,1) \times \frac{SO(5,21)}{SO(5)\times SO(21)}$ of
toroidally compactified heterotic string theory with the following
bosonic field content in five dimensions:
The graviton, 116 scalar 
fields (115 moduli in the matrix $M$ and the dilaton), 26 U(1)
gauge fields and one additional $U(1)$ gauge field, which is the 
dual of the 3-form fieldstrength in five dimensions

We can derive a general T-duality invariant solution  by acting with 
elements of   $SO(5)\times SO(21)$  (the maximal compact subgroup 
of $SO(1,1)\times SO(5,21)$) on the generating solution. 

The electric charges are entries in a vector $\vec Q \in O(5,21)$
with $\vec Q^{T} = ( Q_{1}^{(1)}, 0_{4}; Q_{1}^{(2)} , 0_{20} )$.
Imposing a subset of T-duality transformations, lying in a
coset $SO(5)\times SO(21)/[SO(4)\times SO(20)]$
introduces 24 new charge parameters. Thus the theory is specified by 27 charge
parameters [\ref{hull}]. 
The action of these transformations renders the  
three charges of the generating solution in duality invariant form. Namely 

\begin{eqnarray}
\label{coordinates_inv_5}
  Q_1^{(1)} \to \ \ X &=& 
\frac{1}{2} 
\sqrt{\vec{Q}^{T} {\cal M}_{+} \vec{Q}}, 
                     \ + \ 
\frac{1}{2} 
\sqrt{\vec{Q}^{T} {\cal M}_{-} \vec{Q}},
\nonumber\\
 Q_2^{(2)} \to \  \ Y & = &
\frac{1}{2} 
\sqrt{\vec{Q}^{T} {\cal M}_{+} \vec{Q}}, 
                     \ - \ 
\frac{1}{2}
\sqrt{\vec{Q}^{T} {\cal M}_{-} \vec{Q}}, 
\end{eqnarray}
while $P$ remains intact. i.e.  it is a singlet
under $T$-duality transformations.

Using these three S- and T-duality invariant ``coordinates''  
enables us to cast
all  parameters in the generating solution into 
S- and T-duality
invariant form. For convenience  we will discuss in the following only
the entropy and the ADM mass. Introducing 
``non-extreme hatted'' quantities we find the entropy
\begin{equation}
\hat X_{i} \equiv\sqrt{ X_{i}^{2}+r_{0}^{2}}, \ \ \   X_{i}=(X,Y,P).
\end{equation} 
\begin{eqnarray}
\label{non_entropy_5}
 S &=& 2 \pi
\left \{ 
\left [
 \hat X \hat Y \hat P + r_{0}^{2} (\hat X + \hat Y + \hat P )
+ \sqrt{ (\hat X^{2} - r_{0}^{2})(\hat Y^{2} - r_{0}^{2})(\hat P^{2} - 
r_{0}^{2})}
- (J_{\phi}-J_{\psi})^{2} \right ]^{1/2} \ + 
\right .
\nonumber\\ & &
\left .
\left [
 \hat X \hat Y \hat P + r_{0}^{2} (\hat X + \hat Y + \hat P )
- \sqrt{ (\hat X^{2} - r_{0}^{2})(\hat Y^{2} - r_{0}^{2})(\hat P^{2} - 
r_{0}^{2})}
- (J_{\phi}+J_{\psi})^{2} \right ]^{1/2}
\right \}.   
\end{eqnarray}
The ADM mass reads
\begin{eqnarray}
\label{mass_5}
M &=&  \hat X + \hat Y + \hat P.
\end{eqnarray}
This is a duality invariant form of the ADM mass in terms of the
non-extremality parameter and the charges. In principle,  the inversion 
of this expression yields  the non-extremality parameter 
in terms of the ADM mass and duality invariant combinations 
of charges.

In the regular BPS limit ($r_{0}\rightarrow 0$,$J_{\phi}=-J_{\psi}$) 
the entropy becomes independent of the moduli and is only a function of the 
bare quantized charges $\beta$ and $\vec{\alpha}$, which are defined in
terms of dressed charges  as: $P = \B/g^{4}$ and $ \vec{Q} \ = \ g^{2}
 M_{{\infty}} \vec{\A}$.
This yields [\ref{cvetic4}]
\begin{eqnarray}
S_{BPS} &=&  2\pi \ \sqrt{(\vec{\A}^{T} L \vec{\A})\ \B \ - \ 4 J_{\psi}^{2} }.
\end{eqnarray}

\subsection{Duality Invariant Black Holes of Toroidally Compactified Type II
String}

The toroidally compactified Type II string theory has $N=8$ 
supersymmetry and a  scalar non-linear $\sigma$-modell
with coset space $\frac{E_{6(6)}}{USp(8)}$. The Neveu-Schwarz-Neveu-Schwarz
(NS-NS) sector of the theory
yields the coset $SO(1,1) \times \frac{SO(5,5)}{SO(5)\times SO(5)}$ 
(which is also  the coset space of the toroidal sector
of the heterotic string).
Since the toroidally compactified heterotic string and the NS-NS
sector of the toroidally compactified Type II string theory have the same
effective action they also have the same classical solutions. Moreover,
it was shown in [\ref{hull}] that the generating solutions for black
holes in toroidally compactified Type II string theory are the
same as the one of toroidally compactified heterotic string, i.e.
the generating solution  of toroidally compactified Type II string 
can be specified with the NS-NS charges, only.
We will use this result here to present general U-duality invariant 
non-extreme solutions of Type II string theory in five dimensions.

The central charge matrix of $N=8$ supergravity in five dimensions 
can be brought to a skew-diagonal
form by the use of an $USp(8)$ transformation [\ref{zumino}]
$Z^{0}= \Lambda Z \Lambda^{T}$. 
In this normal form the central charge matrix has four real eigenvalues
$z_{i}$ ($i = 1,\ldots 4$).
General U-duality invariants can be expressed 
in terms of the central charges as:
\begin{eqnarray}
J_{2n} &\equiv& \mbox{tr} \ (Z^{+} Z)^{2n} \ = \ 2 \ 
\sum_{i = 1}^{4} \ z_{i}^{2n}\ , 
\hspace{2cm} n = 1,2,\ldots.
\end{eqnarray}
Using the $USp(8)$ symplectic matrix $\Omega$ one can introduce  
the cubic invariant [\ref{scherk}] $J_3$  which 
can be expressed in terms of central charge eigenvalues as:
\begin{eqnarray}
J_{3} &\equiv& \mbox{tr} \ (\Omega Z)^{3} \ 
= \ 2 \ \sum_{i = 1}^{4} \ z_{i}^{3},
\end{eqnarray}
and the constraint [\ref{cremmer}]
\begin{eqnarray}
\label{invariants}
J_{1} &\equiv& \mbox{tr} \ (\Omega Z) \ = \ 2 \ \sum_{i = 1}^{4} \ z_{i} = 0.
\end{eqnarray}
Hence, for the generating solution 
the central charge matrix is parametrized by
three independent real eigenvalues 
[\ref{hull}]:
\begin{eqnarray}
z_{1,2} &=& P  \ \pm \ Q_{1} \ \pm \ Q_{2} , \nonumber\\
z_{3,4} &=& -P \ \pm \ Q_{1} \ \mp \ Q_{2}. 
\end{eqnarray}
Since the generating solution is given by three independent charges
the general non-extreme U-duality invariant solution must be a function
of three  U-invariants. 
We choose these three invariants to be
$J_{2},J_{3}$ and $J_{4}$. 
Using the following identities, satisfied by the
generating solution,
\begin{eqnarray}
\label{c1}
  8J_{4}-J_{2}^{2} &=& 256 \  
\left ( 
 P^{2} Q_{1}^{2} \ + \ P^{2} Q_{2}^{2} \ + \ Q_{1}^{2} Q_{2}^{2}  
\right ),
\nonumber\\
\label{c2}
J_{3} &=& 48 \  P \ Q_{1} \ Q_{2},
\nonumber\\
\label{c3}
J_{2} &=& 8Q_{1}^{2}+8Q_{2}^{2} +  8P^{2},  
\end{eqnarray}
enables  us to  solve a cubic equation
  for all three charges of the generating
solution in terms of $U$-invariants ($Q_{1,2},P \rightarrow X_{i}$). 
We find three real solutions
\begin{eqnarray}
\label{general_coor_5}
  X_{i} &=& 
\sqrt{ 
2 \rho^{1/3} \ \mbox{cos}(\frac{\varphi+ 2 n_{i} \pi }{3}) +
 \frac{1}{24} J_{2}
}, 
\hspace{2cm} n_{1,2,3} = 0,1,2
\end{eqnarray}
with
\begin{eqnarray}
\rho  &=& \frac{1}{144} \sqrt{ 7 J_{2}^{2} - 24 J_{4} },
\hspace{2cm}
\rho \ \mbox{cos} \ \varphi \ = \ 
\frac{1}{1024} J_{2} \left ( 
                    \frac{17}{108} J_{2}^{2} - \frac{2}{3}J_{4} 
                    \right )
 + \frac{1}{2}\left ( \frac{J_{3}}{48} \right )^{2} .
\end{eqnarray}
The entropy and the ADM mass 
for general non-extreme U-duality invariant black holes
in five dimensions is (\ref{non_entropy_5})  and  (\ref{mass_5})
in these coordinates\footnote{
A special case of this general non-extreme solution is 
a Kerr-Newman type solution with $Q_{2}=Q_{1}=P$.
This solution can be completely determined by $J_{3}$.
In this special case the non-extremality parameter 
is  $ r_{0}^{2} = \left ( \frac{M}{3} \right )^{2} -  
  \left ( \frac{1}{48} J_{3} \right )^{1/3}$
and the BPS mass reads
$M_{BPS}^{2} = \left ( \frac{9}{16} J_{3} \right )^{2/3}$.
In the extremal limit $r_{0}=2(a_{1}+a_{2})^{2}$ 
with $a_{1,2} \geq 0$ we can eliminate one angular momentum 
parameter by keeping the other one and the non-extremality parameter. We find
$J_{\psi} + J_{\phi} = \frac{1}{2} \ ( \hat P_{+} - \hat P_{-} )$
with $ \hat P_{\pm}=(\hat P \pm r_{0})^{3/2}$. 
The entropy in the extremal limit reads
$ S_{EXT} = \pi
\left \{ 
\sqrt{ 2 (\hat P_{+} + \hat P_{-})^{2} -
          (\hat P_{+} - \hat P_{-} - 4 J_{\psi})^{2}} 
\  + \ \sqrt{\hat P_{+} - \hat P_{-}}
\right \}$.
An analogous discussion holds for the toroidally compactified
heterotic string in five dimensions (with $N=4$ supersymmetry).
}.
In the regular BPS limit the entropy is only a function of $J_{3}$ 
[\ref{ferrara},\ref{hull}] and one angular momentum $J_\psi=-J_\phi\equiv J$

\begin{eqnarray}
  S_{BPS} &=&  {\pi} \ \sqrt{ \textstyle{{J_{3}}\over {12}} - 8J^{2} }.
\end{eqnarray}

\section{N = 4 Black Holes in Four Dimensions}
\resetcounter

In $D=4$ the generating solution 
is parametrized by {\em five} charges 
[\ref{tseytlin},\ref{hull}].
Although we will comment on these general non-extreme black holes
at the end of this chapter, we were not able to find the
general non-extreme duality invariant thermodynamic quantities
within our approach. However, starting with a truncated version of the
generating solution,  specified by {\em four}-charges,
and further reducing it to two with ``fixed values'' of moduli, 
we find the  duality invariant combinations of charges specifying
non-extreme solutions of pure $N=4$ supergravity.

In addition, in Section III. we shall address the duality invariant solutions
of torodially compactified string theory. Using the 
results for the non-extreme five (charge) parameter generating solution of 
[\ref{cvetic3}]  and the duality invariant structure of such solutions in
the BPS limit [\ref{tseytlin}], we were able to determine the duality invariant
form of the near-extreme solution.

\subsection{The Generating Solution}

The starting point of the $D=4$  black hole solutions is 
 the Kerr-solution (neutral rotating black hole) 
 \footnote{We take $G_{N}= \frac{1}{4} g^{2} $, $g^{2} = 
e^{\Phi_{\infty}}$.},
The  solution is specified by the 
non-extremality parameter $r_{0}$ and one rotational parameter $a$. 
Kerr metrics are all stationary and axisymmetric with
Killing fields $\xi^{a}=(\frac{\partial}{\partial t})^{a}$
and $\psi^{a}= (\frac{\partial}{\partial \phi})^{a}$.
The event horizons are located where the $r=\mbox{const}$
surface vanishes.
Thus the inner and the outer horizon is given by
\begin{eqnarray}
\label{Kerr_horizons}
  r_{\pm} &=& \frac{r_{0}}{2} \ \pm \ \sqrt{\frac{r_{0}^{2}}{4} \ - \ a^{2} }.
\end{eqnarray} 
provided $r_{0}/2 \geq a$. 
These horizons disappear for $a > r_{0}/2$ giving rise to a 
naked singularity. The limit $a \rightarrow r_{0}/2$ is known as
the extremal limit. 
The physical singularity of the vacuum Kerr solution is time-like and
concentrated on a ring at $r=0$ and $\theta=\pi/2$. 

Performing four boost transformations on the Kerr solution, i.e.
symmetry transformations  of stationary solutions, generate  
charged  solutions (for more details see
 [\ref{cvetic3}]) specified by the four charges:

\begin{eqnarray}
\label{charge_def}
  P_{1}^{(1,2)} &\equiv& P_{1,2} \ = \ 2 \ r_{0} \ \mbox{sinh}\delta_{p1,2} \
                             \mbox{cosh}\delta_{p1,2},
\nonumber\\
 Q_{2}^{(1,2)}  &\equiv& Q_{1,2} \ = \ 2 \ r_{0} \ \mbox{sinh}\delta_{q1,2} \
                            \mbox{cosh}\delta_{q1,2}.
\end{eqnarray} 
Here 
$(P_1^{(1)}, \ P_1^{(2)})$ [($Q_2^{(1)}, \ Q_2^{(2)}$)] are the  magnetic
[electric] charges 
associated with  $U(1)$ gauge fields 
$A_{\mu\, 1}^{(1,2)}$ [$A_{\mu\, 2}^{(1,2)}$] in the (momentum, winding) sector 
 of the string theory, associated with the first [second] compactified
 direction.
These charges are entries in a vector $\vec Q \in O(6,22)$
and $\vec P \in O(6,22)$
with $\vec Q^{T} = ( 0 , Q_{2}^{(1)}, 0_{4}; 0,  Q_{2}^{(2)} , 0_{20} )$
and $\vec P^{T} = ( P_{1}^{(1)}  , 0_{5}; P_{1}^{(2)} , 0_{21} )$.

The corresponding  charged  rotating solution  has
the angular momentum specified
as [\ref{cvetic2}] 
\begin{eqnarray}
\label{angular_def}
  J  &=& 4 \ r_{0} \ a  \left ( 
                     \prod_{i=1}^{4}  \mbox{cosh}\delta_{i} \ - \
                     \prod_{i=1}^{4}  \mbox{sinh}\delta_{i}
                   \right ).
\end{eqnarray} 
This generating solution is thus  parametrized by  the non-extremality
parameter
$r_{0}$ and the four charges (\ref{charge_def}) and the angualr momentum
(\ref{angular_def}). 

The four dimensional metric and the fields for this
solution  have been given in [\ref{cvetic3}]. The  entropy and 
the ADM mass of this charged  rotating solution is 
[\ref{cvetic2}]:\footnote{and the dilaton charge is of the form:
$\Sigma  =    {\hat P}_{1} +  {\hat P}_{2} 
                - {\hat Q}_{1}  -  {\hat Q}_{2}$.} 
\begin{eqnarray}
\label{non_area}
 S &=&
 \frac{\pi}{r_{0} } \  
\left \{ \ r_{+} \ 
    \prod_{i=1}^{4}  \sqrt{ \hat Q_{i} + r_{0} }
 \ + \ r_{-} \
    \prod_{i=1}^{4}  \sqrt{ \hat Q_{i} - r_{0} } 
\right \},  
\end{eqnarray}
\begin{eqnarray}
\label{non_mass_general}
    M  &=& 2r_{0}  \ \sum_{i=1}^{4} \ 
                 \ ( \mbox{cosh}^{2}\delta_{i} \ +\sinh^2\delta_i )
                        \ = \ {\hat Q}_{1} \ + \ {\hat Q}_{2} \ + \
              {\hat P}_{1} \ + \ {\hat P}_{2}. 
\end{eqnarray}

Note that the above  solution, specified by four-charges
(\ref{charge_def}),  corresponds to  a special case of
 the generating solution for torodially compactified  heterotic and Type II
[\ref{hull}] string, which 
is specified by {\it five charge parameters}.
\subsection{Duality Invariant Black Holes  in N = 4 Supergravity}

We start with this four-charge parameter generating solution of toroidally
compactified heterotic (or Type II) string  (described in the
previous Section)  in order to  obtain the general  S-duality invariant
solutions of pure $N=4$ supergravity theory (i.e., with ``fixed moduli'').

The scalar sector  of $D=4$  torodially compactified heterotic 
string  corresponds to 
the  non-linear $\sigma$-model with coset space
$\frac{SU(1,1)}{U(1)} \times \frac{O(6,22)}{O(6)\times O(22)}$. 
We can  now derive a  T-duality invariant solution by acting with
elements  of $O(6)\times O(22)$ (maximal compact subgroup 
of $O(6,22)$) on the  four-charge parameter generating solution.
This solution is  thus specified by 54 independent 
charge parameters (i.e. 4 charge parameters of the generating solution and 50
parameters of the  
coset $[SO(6)\times SO(22)]/[SO(4)\times SO(20)]\subset O(6)\times O(22)$).
Namely, the solution is specified by 28 electric and 28 magnetic charges 
subject to {\it two} charge constraints.

Thus, the  action of the transformations renders the
four charges of the generating solution to a T-duality invariant form:
\begin{eqnarray}
\label{coordinates}
 P_{1,2} &\rightarrow& \frac{1}{2} \sqrt{\vec{P}^{T} {\cal M}_{+} \vec{P}} 
                     \ \pm \ 
                     \frac{1}{2} \sqrt{\vec{P}^{T} {\cal M}_{-} \vec{P}},
\nonumber\\
 Q_{1,2} &\rightarrow& \frac{1}{2} \sqrt{\vec{Q}^{T} {\cal M}_{+} \vec{Q}} 
                     \ \pm \ 
                     \frac{1}{2} \sqrt{\vec{Q}^{T} {\cal M}_{-} \vec{Q}}.
\end{eqnarray}
The corresponding solution obeys two $O(6,22)$ invariant constraints on the
charges:
\begin{eqnarray}
\label{ab_constraints}
 \vec{P}^{T} {\cal M}_{+} \vec{Q} &=& 0,  
\hspace{2cm}
 \vec{P}^{T} LM_{\infty}L \vec{Q} \ = \ 0
\end{eqnarray}
and is not invariant under $SL(2,{\bf R})$ transformations.
The dressed charges are related to 
bare  quantized  charges in the following way:
\begin{eqnarray}
\label{charge_relation1}
 \vec{P} &=& L \ \vec{\B},
\hspace{2cm}
 \vec{Q} \ = \ e^{\Phi_{\infty}} M_{{\infty}}
           (\vec{\A} + \Psi_{\infty} \vec{\B}). 
\end{eqnarray}

Setting, 
$\Phi_{\infty} = \Psi_{\infty} = 0$ and $M_{\infty}=I_{28}$, allows one
to perform an $SO(2)$ (maximal compact subgroup of S-duality group
$SL(2, {\bf R})$) transformation specified by angle $\theta$ , which  
enables one
to  find a manifestly S-duality
invariant form of the  solution. This transformation removes one of the 
 charge
constraints. However, one constraint on charges remains and can be cast
in the form:
\begin{eqnarray}
\label{angle_constraints}
 \mbox{tan} \ \theta  &=& - \frac{1}{2 \vec{\A}^{T} \mu_{+} \vec{\B}}
                       \left(
                              g^{2} \ \vec{\A}^{T} \mu_{+} \vec{\A}
                        \ - \ \frac{1}{g^{2}} \ \vec{\B}^{T} \mu_{+} \vec{\B}
                        \ - \ \gamma(\mu_{+})  
                       \right)
\nonumber\\ &=&
     - \frac{1}{2 \vec{\A}^{T} L \vec{\B}}
                       \left(
                              g^{2} \ \vec{\A}^{T} L \vec{\A}
                        \ - \ \frac{1}{g^{2}} \ \vec{\B}^{T} L \vec{\B}
                        \ - \ \gamma(L)  
                       \right)
\end{eqnarray}
with 
\begin{eqnarray}
\label{gamma_def}
 \gamma(y)  &=& \sqrt{4 \ (\vec{\A}^{T} y  \vec{\B})^{2}
                     \ + \
                     (g^{2} \ \vec{\A}^{T} y \vec{\A}
                       \ - \ \frac{1}{g^{2}} \ \vec{\B}^{T} y \vec{\B})^{2} }.
\end{eqnarray}
This constraint can be solved to give 
\begin{eqnarray}
\mu_{+}\vec\A &\equiv& (I + L)\vec\A \ = \ 2L \vec\A,
\hspace{2cm}
\mu_{+}\vec\B \ \equiv\ (I + L)\vec\B \ = \ 2L \vec\B.  
\end{eqnarray}
Thus we find $\mu_{-}\vec\A=\mu_{-}\vec\B=0$ and recover the result that
a consistent $N=4$ truncation of the heterotic string theory
in four dimensions removes the ``left-movers'' by identifying  
the momentum with the winding modes and sets, in addition, 
the 16 Yang-Mills gauge fields of the heterotic string theory to zero
\footnote{Thus the notation is somehow
redundant: $\vec{\A}^{T} L \vec{\A} = 2 \sum_{i=1}^{6} \A_{i}\A^{i}$.
The analogous notation holds for the bare magnetic charges as well.
The  surviving symmetry group is only $SO(6)$, but for convenience we keep the
``trivial'' $SO(6,6)$ notation.}. 
As the last step  one can transform the solution
back to the physical, dressed charges  with arbitrary 
asymptotic values of the dilaton and axion fields.

Thus, in order to ensure that the  remaining charge constraint 
(\ref{angle_constraints}) is  (automatically) satisfied we
ended up with an S-duality invariant solution of  
pure $N=4$ supergravity (with ``fixed moduli'')  and the 
coset space $\frac{SU(1,1)}{U(1)}$. Namely,
the  corresponding matrix of the moduli 
is diagonal and constant
and only the dilaton
and the axion remain as dynamical scalar degrees of freedom.
This truncated model contains now 6 vector fields and
parametrizes 12 charges.

The same solution
can be obtained directly  by starting with the  special 
case of the generating solution  with 
$P_{1}=P_{2}$ and $Q_{1}=Q_{2}$.
The corresponding solution is
specified by two harmonic functions and fits into the
pointlike supersymmetric Wilson-Israel-Perj\'{e}s solutions
of [\ref{ortin}]. There the non-extreme entropy was also given
in a duality invariant form and the corresponding thermodynamic
quantities can be read off straightforward.

The charges of the generating solution $P_{1}$ and $Q_{1}$ 
emerge in the general solution of pure $N=4$ supergravity
to two (S-duality invariant) coordinates $X$ and $Y$ respectively.
\begin{eqnarray}
\label{coordinates_inv}
 Q_1=Q_2\to {X} &=& 
\sqrt{\frac{1}{2} F(L,-\Gamma)  },
\hspace{2cm}
P_1=P_2\to {Y} \ = \ 
\sqrt{ \frac{1}{2} F(L,\Gamma)  },  
\end{eqnarray}
with
\begin{eqnarray}
 F(L,\pm\Gamma)  &=& 
   \frac{1}{2g^{2}} \ 
\left ( 
 \vec{Q}^{T} L \vec{Q} \ + \
 \vec{P}^{T} L \vec{P} \ \pm \  
 \Gamma(L)
\right ),
\nonumber\\
\label{Gamma_def2}
 \Gamma(L)  &=& \sqrt{4 \ (\vec{P}^{T} L M_{\infty}L \vec{Q})^{2}
                     \ + \
                     (\vec{Q}^{T} L \vec{Q}
                       \ - \ \vec{P}^{T} L \vec{P})^{2} }.
\end{eqnarray}
These coordinates are - analogous to our five dimensional considerations - 
invariant under $SO(6)$ and $SL(2,{\bf R})$
transformations\footnote{These coordinates are also
invariant under general $O(6,22)$ transformations.}. 
Thus they provide a basis to discuss T- and S-duality
invariant quantities.  In terms of the ``non-extreme hatted'' quantities
\begin{equation}
\hat X \equiv\sqrt{ X^2+r_{0}^{2}}, \ \ \ \ \ \ 
\hat Y\equiv\sqrt{ Y^2+r_{0}^{2}}
\end{equation} 
the ADM mass and the dilaton charge are of the form
\begin{eqnarray}
\label{ADM_mass_non_extreme}
  M &=& 2 \ \hat X \ + \ 2 \ \hat Y,
\hspace{2cm}
 \Sigma  \ = \ g^{4} ( \ 2 \ \hat X \ -  \ 2 \ \hat Y \ ).
\end{eqnarray}
The non-extremality parameter $r_0$, which is
invariant under T- and S-duality,  can in turn  be determined in
terms of the  duality invariant combinations of charges $X$ and $Y$ and the 
ADM mass as:
\begin{eqnarray}
\label{non_extreme_parameter}
  4 r_{0}^{2} &=&\frac{M^{2}}{4} \ - \ F(L,\Gamma) \ - \ F(L,-\Gamma)
   \ + \ \left ( \frac{ F(L,\Gamma)- F(L,-\Gamma)}{M} \right )^{2}
\end{eqnarray}
This gives then the corresponding solution for the entropy of the
Kerr solution with non-vanishing total angular momentum $J=a M$
\begin{eqnarray}
\label{kerr_entropy}
 S &=&
\frac{\pi}{r_{0}}   
\left \{ r_{+} ( \hat X + r_{0}) ( \hat Y + r_{0})
  +  r_{-} ( \hat X - r_{0}) ( \hat Y - r_{0})
\right \}.   
\end{eqnarray}

To calculate the surface gravity 
$\kappa = \lim_{r \rightarrow r_{+}} \ \sqrt{g^{rr}} \ \partial_{r}
           \ \sqrt{-g_{tt}}_{| \theta=0}$
of the Kerr solution we need the metric components of the
solution in duality invariant form. Using the results above
they can be easily read off of the generating solution in
[\ref{cvetic2}].
We find the surface gravity to be
\begin{eqnarray}
\kappa &=& 2 \ r_{0} \ 
\frac{ r_{+}-r_{-}}{ r_{+} ( \hat X + r_{0}) ( \hat Y + r_{0})
  +  r_{-} ( \hat X - r_{0}) ( \hat Y - r_{0})}.
\end{eqnarray}
The Hawking temperature of the black hole
is $T = \frac{\kappa}{2\pi}$.
Hence we recover the established relation  [\ref{proeyen}, \ref{sen1}]:
$ T S = r_{+} - r_{-} $. 
\\
Moreover,
the Killing field $\chi^{a}= \xi^{a} + \Omega \psi^{a}$ is defined
to vanish at the horizon ($r=r_{+}$). This yields the angular velocity
\begin{eqnarray}
\Omega &=& 4 \ r_{0}^{2} \ 
           \frac{a}{ r_{+} ( \hat X + r_{0}) ( \hat Y + r_{0})
                  +  r_{-} ( \hat X - r_{0}) ( \hat Y - r_{0}) }.
\end{eqnarray}

\subsection{Extremal and BPS Limits}

The extremal limit, in which the event horizon is about to disappear,
is given by

\begin{eqnarray}
\label{extr_mass}
  M^{2} &\rightarrow& 2  F(L,\Gamma)  \ + \
                      2  F(L,-\Gamma) \ + \ 
             4 \sqrt{ F(L,-\Gamma) F(L,\Gamma) + 4 J^{2}}.
\end{eqnarray}
The extremal ADM mass (\ref{extr_mass}) reduces to the BPS mass [\ref{cvetic1}]
in the limit $J \rightarrow 0$. 
This regular BPS limit, which is a special extremal limit,
has no naked singularity. In the extremal limit we have $r_{0} \rightarrow 0$
if and only if  $J \rightarrow 0$. Hence the inequality
$r_{0}^{2} \geq 4J^{2}$ is saturated in the regular BPS limit.
Thus the regular BPS limit is the extreme limit with vanishing
angular momentum and the rotating non-extreme black holes 
lose all their angular momentum on their way to extremality.
A possible dynamical process, that drives black holes to extremality, while
they lose all their angular momentum, is the
``Penrose Process'' [\ref{penrose}].
Once the BPS limit is reached due to this process,
the remaining black hole is spherically symmetric. 
It is rather interesting that this classical 
process is compatible with the Bogomol'nyi-Gibbons-Hull bound,
since the process 
stops if an irreducible mass is reached [\ref{christodoulou}].

The Bogomol'nyi-Gibbons-Hull bound itself [\ref{gibbons}]
determines the mass of particle states in
extended supersymmetric theories in terms of the central
charge matrix $Z_{AB}$. 
(For a review see [\ref{auria}] and reference therein.)

\begin{eqnarray}
 M  &\geq& \mbox{max} \ |Z_{AB}|.
\end{eqnarray}
The central charge matrix $Z_{AB}$ can be brought to a skew-diagonal
form by a $SU(4)$ transformation [\ref{zumino}]: $ Z^{0}_{AB} = \Lambda  Z_{AB} \Lambda^{T} $.
In terms of duality invariant quantities the absolute values
of these eigenvalues are [\ref{duff}]
\begin{eqnarray}
\label{eigenvalues}
|z_{1,2}| &=& \sqrt{2F(L,\Gamma)} \ \pm \ \sqrt{2F(L,-\Gamma)}.
\end{eqnarray}
In the regular BPS limit we have
\begin{eqnarray}
\lim_{r_{0}\rightarrow 0} M &=& M_{BPS} \ = \ |z_{1}|,
\hspace{2cm}
\lim_{r_{0}\rightarrow 0} \Sigma \ = \ \Sigma_{BPS} \ = \ -|z_{2}| g^{4}.
\end{eqnarray}
Moreover, the moduli fields, the dilaton and the axion
are {\em fixed} values in the BPS limit at the horizon.
The corresponding entropy only depends on
the  bare quantized charges [\ref{larsen}].
Furthermore it has been shown in [\ref{ferrara}] that this result
can be understood from a point of view of ``supersymmetric attractors''.
In $N=4$ supergravity we only have to find the fixed values of 
the dilaton and the axion at the horizon. 
Following [\ref{ferrara}]
the dilaton charge has to vanish in the BPS limit ($z_{2}=0$).
This implies $\Gamma(L)=0$ and yields

\begin{eqnarray}
  \Psi_{fix} &=&  - \ \frac{\vec{\A}^{T} L \vec{\B}}{\vec{\B}^{T} L \vec{\B}},
\hspace{2cm}
g^{4}_{fix} \ = \ \frac{(\vec{\B}^{T} L \vec{\B})^{2}}
                       {(\vec{\A}^{T} L \vec{\A})(\vec{\B}^{T} L \vec{\B})
                         - (\vec{\A}^{T} L \vec{\B})^{2}}.
\end{eqnarray}
The corresponding area of the horizon is proportional to the largest
eigenvalue of the central charge at the fixed points of the dilaton
and the axion.
\begin{eqnarray}
 |z_{1}|_{fix}
      &=& 2 \ \sqrt{2} \
            \left [ 
              (\vec{\A}^{T} L \vec{\A})(\vec{\B}^{T} L \vec{\B})
               - (\vec{\A}^{T} L \vec{\B})^{2} 
            \right ]^{1/4}.
\end{eqnarray}
The remaining theory in the BPS limit has $N=1$ supersymmetry and
the Bekenstein-Hawking entropy is  
\begin{eqnarray}
S_{BPS} &=& \frac{\pi}{16} \  |z_{1}|_{fix}^{2}.
\end{eqnarray}

\subsection{Duality Invariant Black Holes in N = 4 String Theory}
Now we turn to a discussion of  general static case of toroidally compactified
heterotic string  with the scalar coset space
$\frac{SU(1,1)}{U(1)} \times \frac{O(6,22)}{O(6)\times O(22)}$.
Although we were unable to present the complete duality invariant solution,
however, we found  the  T- and S-duality  invariant entropy of  general {\it
near-extremal} solutions.
 
 Starting with the non-extreme
five-paramter generating solution of [\ref{tseytlin},\ref{cvetic3}]
the general $N=4$ spherically symmetric solution has 
the time component of the metric, which can be written in the form:
$g_{tt}=\pi (r+r_{0})(r-r_{0})S^{-1}(r)$. 
The function  $S(r)$, which  can be written in the form
\begin{eqnarray}
S(r) &=& \pi \ \prod_{i=1}^{4} \sqrt{(r +\lambda_{i})},
\end{eqnarray}
 specifies at the horizon ($r=r_{0}$) the entropy of the 
 non-extreme  solution, i.e.
$S \equiv S(r_{0})$. 

For the five-parameter generating solution 
$\lambda_{i}$ depend  on five-charge parameters, and the
non-extremality parameter $r_0$, which enters the expressions 
for $\lambda_i$
as a function of $r_{0}^{2}$, only.

The general solution, 
parametrized by 28 electric
and 28 magnetic charges $Q_{i}$ and $P_{i}$ ($i=1 \ldots 28$) 
respectively, is obtained by imposing on the generating
solution $[SO(6) \times SO(22)]/[SO(4) \times SO(20)] \subset O(6,22)$
and $SO(2) \subset SL(2,{\bf R})$ transformations [\ref{cvetic3}].

The outstanding problem is to find the duality invariant form of
the parameters $\lambda_i$\footnote{U-duality invariant structure of 
non-extreme black holes  have also been considered in [\ref{rajaraman}]. 
However,  in that case we were 
unable  to reproduce the BPS limit of the  {\it five-charge parameter}
generating solution.} with
$\lambda_{i} \equiv \lambda_{i}(r_{0}^{2},Q_{i},P_{i})$.    
However, since $\lambda_i$  depend on $r_{0}^{2}$ only, we can
determine the T- and S-duality invariant form of the entropy in the 
near-extremal limit.The entropy reads
\begin{eqnarray}
\label{proposal}
S &=& \pi \ \prod_{i=1}^{4} \sqrt{\lambda_{i}^{[0]}} \ + \ 
\frac{\pi}{2} \ r_{0} \ \sum_{i<j<k} 
\lambda_{i}^{[0]} \lambda_{j}^{[0]} \lambda_{k}^{[0]}
\ + \ {\cal O}(r_{0}^{2}) 
\end{eqnarray}
Here the $\lambda_{i}^{[0]}$ are the eigenvalues in the BPS limit.
Starting with the BPS five-charge (generating) solution of [\ref{tseytlin}]
and imposing T- and S-duality transformations one finds for
the general 56-charge configuration the following T- and S-duality
invariant quantities

\begin{eqnarray}
\prod_{i=1}^{4} \lambda_{i}^{[0]} &\equiv& S^{2}_{BPS}/\pi
\ = \  \frac{1}{4} \ F(L,\Gamma) F(L,-\Gamma),
\nonumber\\
\sum_{i=1}^{4} \lambda_{i}^{[0]}  &\equiv& M_{BPS}
\ = \  \sqrt{ F({\cal M}_{+},\Gamma) } + \sqrt{ F({\cal M}_{+},-\Gamma) },
\nonumber\\
\sum_{i<j} \lambda_{i}^{[0]} \lambda_{j}^{[0]}  &=&
\frac{1}{2g^{2}} 
\left( 
 \vec{Q}^{T} L \vec{Q} + \vec{P}^{T} L \vec{P}
\right)
+ \sqrt{ F({\cal M}_{+},\Gamma) F({\cal M}_{+},-\Gamma)},
\nonumber\\
\sum_{i<j<k} \lambda_{i}^{[0]} \lambda_{j}^{[0]} \lambda_{k}^{[0]}  &=& 
\frac{1}{4g^{2}M_{BPS}}
\left \{
 M_{BPS}^{2} 
\left (
 \vec{Q}^{T} L \vec{Q} + \vec{P}^{T} L \vec{P} 
\right )
\right .
\nonumber\\ & &
\left .
\ \ \ \ \ \ \ \ \ \ \ \ \ \ -  
 (\vec{Q}^{T} {\cal M}_{+} \vec{Q} - \vec{P}^{T} {\cal M}_{+} \vec{P})
 (\vec{Q}^{T} L \vec{Q} - \vec{P}^{T} L \vec{P})
\right .
\nonumber\\ & &
\left .
\ \ \ \ \ \ \ \ \ \ \ \ \ \ - 
4 (\vec{Q}^{T} LM_{\infty}L \vec{P}) (\vec{Q}^{T} {\cal M}_{+} \vec{P}) 
\right \}.
\end{eqnarray}
Here we used the following definitions
\begin{eqnarray}
 F({\cal M}_{+},\pm\Gamma)  &=& 
   \frac{1}{2g^{2}} \ 
\left ( 
 \vec{Q}^{T} {\cal M}_{+} \vec{Q} \ + \
 \vec{P}^{T} {\cal M}_{+} \vec{P} \ \pm \  
 \Gamma({\cal M}_{+})
\right )
\nonumber\\
 \Gamma({\cal M}_{+})  &=& \sqrt{4 \ (\vec{P}^{T} {\cal M}_{+} \vec{Q})^{2}
                     \ + \
                     (\vec{Q}^{T} {\cal M}_{+} \vec{Q}
                     \ - \ \vec{P}^{T} {\cal M}_{+} \vec{P})^{2} }.
\end{eqnarray}

Since the five-charge parameter solution  is the generating solution for 
$D=4$ black holes of toroidally compactified Type II string, the U-duality
 invariant  form of  {\it near-extremal} solutions could be
obtained along similar lines.

On the other hand  the  structure of 
 T- and S-duality [U-duality] invariant solutions for {\it non-extremal}
 solutions of toroidally compactified heterotic [Type II] string  theory in $D=4$
 remains an open problem.


\vspace{0.3cm}

{\bf Acknowledgement:} We would like to thank F. Larsen for many helpful
  discussions and participation in the early stage of this work. The work 
  is supported by U.S. DOE Grant Nos. DOE-EY-76-02-3071, the National 
Science Foundation Career Advancement Award No. PHY95-12732 and the NATO
collaborative research grant CGR No. 940870 (M.C.).

\vspace{0.3cm}

\section{Appendix}
\resetcounter

In this appendix we summarize some basic facts concerning 
low-energy effective actions and duality symmetries of heterotic 
and Type II string theory. For a comprehensive review we refer
to [\ref{sen2}].

\subsection{Heterotic Superstring on $T^{6}$ (in D = 4)}

The moduli fields of the heterotic string theory in four dimensions
can be combined to an $O(6,22)$ matrix valued scalar field 
$M$ [\ref{sen2}] with
\begin{eqnarray}
 M &=& M^{T}, 
\hspace{2cm} MLM^{T} \ = \ L.
\end{eqnarray}
The gauge sector in four dimensions consists out of 28 U(1) gauge fields
$A_{\mu}^{(a)}$ ($a=1,\ldots 28$). The antisymmetric tensor and
the dilaton combine in four dimensions to a complex scalar $\lambda$
with the help of the Poincar{\'e} duality, which transforms on-shell 
the antisymmetric tensor $B_{\mu\nu}$ to an axion $\Psi$.
Since we consider SL(2,{\bf R}) invariance,
which is - first of all - only a symmetry of the equations of motion, 
we restrict ourselves to the on-shell case.
This corresponding on-shell action is invariant under  $O(6,22)$ and
$SL(2,{\bf R})$ transformations. The subgroup
$O(6,22;{\bf Z})$ (T-duality group) is an exact symmetry of the 
string theory, whereas the subgroup $SL(2,{\bf Z})$ (S-duality group)
is a conjectured non-perturbative symmetry of string and
{\em M-}theory. 

\subsubsection{ $O(6,22)$ Transformations}

$\Omega \in O(6,22)$ if $\Omega^{T}L\Omega = L$ with

\begin{eqnarray}
L &=& 
\left (
   \begin{array}{ccc}
   0     & I_{6} & 0       \\
   I_{6} & 0     & 0       \\
   0     & 0     & I_{16}  \\
\end{array}
\right )
\end{eqnarray}
The corresponding fields transform under $O(6,22)$ transformtions
as follows
\begin{eqnarray}
M  \ \rightarrow \ \Omega \ M \ \Omega^{T},
\hspace{2cm} 
A_{\mu}^{(a)}  \ \rightarrow \ \Omega_{ab} \ A_{\mu}^{(b)}, 
\end{eqnarray}
whereas  $g_{\mu\nu}$,$B_{\mu\nu}$ and $\Phi$ are invariant under
$O(6,22)$ transformtions
The projection operators we use are
$\mu_{\pm}= M_{\infty} \pm L$, ${\cal M}_{\pm} =(LML)_{\infty} \pm L$.

\subsubsection{ $SL(2,{\bf R})$ Transformations}

In general the axion-dilaton and the gauge fields transform under 
$SL(2,{\bf R})$ transformations, while the
metric $g_{\mu\nu}$ and the moduli $M$ are invariant. 
The corresponding SL(2,{\bf R}) transformations of the dressed 
physical charges  and the string coupling constant are:

\begin{eqnarray}
\label{sl_phy_ch}
{\vec Q} &\rightarrow& (c \ \Psi_{\infty} \ + \ d) \ {\vec Q}
                       \ + \
                       c \ e^{- \Phi_{\infty}} \ M_{\infty} \ L 
                       \ {\vec P}, 
\nonumber\\   
{\vec P} &\rightarrow& (c \ \Psi_{\infty} \ + \ d) \ {\vec P}
                       \ - \
                       c \ e^{- \Phi_{\infty}} \ M_{\infty} \ L 
                       \ {\vec Q}, 
\nonumber\\ 
 g^{2} &\rightarrow& [(c \Psi_{\infty} +  d)^{2}  
                       \ + \
                       c^{2} e^{- 2 \Phi_{\infty}}]^{2} \ g^{2}.
\end{eqnarray}
The corresponding SL(2,{\bf R}) transformations of the bare charges
are 
\begin{eqnarray}
\label{S_duality_undr}
  \left (
   \begin{array}{c}
   \vec{\A}  \\
   \vec{\B}  \\
\end{array}
\right )
&\rightarrow&
\lagrange \omega \lagrange^{T}
 \left (
   \begin{array}{c}
   \vec{\A}  \\
   \vec{\B}  \\
\end{array}
\right )
\ = \
\left (
   \begin{array}{cc}
   a & -b  \\
   -c & d  \\
\end{array}
\right ) \
\left (
   \begin{array}{c}
   \vec{\A}  \\
   \vec{\B}  \\
\end{array}
\right )
\end{eqnarray}
with
\begin{eqnarray}
\label{S_duality_def}
\lagrange &=& 
\left (
   \begin{array}{cc}
   0 & 1   \\
   -1 & 0  \\
\end{array}
\right ),
\hspace{2cm}
\omega \ = \
\left (
   \begin{array}{cc}
   d & c  \\
   b & a  \\
\end{array}
\right ),
\hspace{1cm} \ 
ad-bc \ = \ 1 
\end{eqnarray}
Note that an $SO(2) \subset SL(2,R)$ transformation
preserves the special background with $\Psi = \Phi = 0$ [\ref{sen1}].
This $SO(2)$ transformation can be parametrised by one angle $\theta$:
\begin{eqnarray}
\label{S1}
  \omega
&=&
\left (
   \begin{array}{cc}
   \mbox{cos} \theta  & \mbox{sin}\theta   \\
   - \mbox{sin}\theta & \mbox{cos}\theta   \\
\end{array}
\right ) \
\end{eqnarray}
Using this $SO(2)$ transformations one finds 
\begin{eqnarray}
\label{trafo_laws}
 g^{2} \ \vec{\A}^{T} y \vec{\A} &\rightarrow&   
                                 g^{2} \ \vec{\A}^{T} y \vec{\A} \ + \
                                 \mbox{tan} \ \theta \ \vec{\A}^{T} y \vec{\B},
\nonumber\\
\frac{1}{g^{2}} \ \vec{\B}^{T} y \vec{\B} &\rightarrow&  
                                \frac{1}{g^{2}} \ \vec{\B}^{T} y \vec{\B} \ - \
                                \mbox{tan} \ \theta \ \vec{\A}^{T} y \vec{\B}.
\end{eqnarray}

\subsection{The Heterotic/Type II String  on $T^{5}$ (in D = 5)}

For the low energy effective heterotic string compactified on $T^{5}$
[\ref{cvetic4}] the T-duality group is $SO(5,21,{\bf Z})$
with analogous transformations of the fields with respect to the
four dimensional case. On the other hand, the S-duality group becomes 
simply $SO(1,1,{\bf Z})$.

The low energy effective Type II 
string compactified on $T^{5}$
[\ref{hull}, \ref{scherk}] has T-duality group  
$SO(5,5,{\bf Z})$ with analogous 
transformations of the fields with respect to the
four dimensional case. Moreover,  
the U-duality group in five dimensions is $E_{6(6)}(\bf Z)$ 
[\ref{townsend}].
The maximal compact subgroup is the unitary symplectic
group $USp(8)$. The 27 abelian gauge fields 
of $N=8$ supergravity and their corresponding
charges transform as a $\bf 27$ vector of $E_{6(6)}$.

%
%

\newpage

\begin{center}
{\bf References}
\end{center}

\begin{enumerate}
\item
\label{horowitz}
G. Horowitz, UCSBTH-96-07, gr-qc/9604051.
\item
\label{maldacena}
J. Maldacena, Ph.D. thesis, hep-th/9607235.
\item 
\label{cveticR}
M. Cveti{\v{c}}, UPR-714-T, hep-th/9701152.
\item
\label{luest}
K. Behrndt, R. Kallosh, J. Rahmfeld, M. Shmakova and W.K. Wong,
 Phys. Rev. {\bf D54} (1996) 6293;
\\
G. Lopes-Cardoso, D. L{\"u}st and T. Mohaupt, 
Phys. Lett. {\bf B388} (1996) 266;
\\
K. Behrndt, G. Lopes-Cardoso, B. de Wit, R. Kallosh, D. L{\"u}st and 
T. Mohaupt, hep-th/9610105;
\\
W. Sabra, hep-th/9703101.
\item
\label{sen1}
A. Sen, Nucl. Phys.  {\bf B440 } (1995) 421;
\\
A. Sen, Mod. Phys. Lett. {\bf A10} (1995) 2081.
\item
\label{proeyen}
R. Kallosh, A. Linde, T. Ortin, A. Peet and A. Van Proeyen, 
Phys. Rev. {\bf D46 } (1992) 5278.
\item
\label{ferrara}
S. Ferrara and R. Kallosh,  
Phys. Rev. {\bf D54} (1996) 1514,
Phys. Rev. {\bf D54} (1996) 1525.
\item
\label{ortin}
T. Ortin, Phys. Rev. {\bf D47 } (1993) 3136. \\
R. Kallosh and T. Ortin, Phys. Rev. {\bf D48} (1993) 742. \\
E. Bergshoeff, R. Kallosh and T. Ortin, Nucl. Phys.  {\bf B478 } (1996) 156.
\item
\label{rajaraman}
R. Kallosh and A. Rajaraman, 
Phys. Rev. {\bf D54} (1996) 6381.
\item
\label{cvetic1}
M. Cveti{\v{c}} and D. Youm,
Phys. Rev. {\bf D53 } (1996) 584.
\item
\label{cvetic2}
M. Cveti{\v{c}} and D. Youm, Phys. Rev. {\bf D54 } (1996) 2612.
\item
\label{cvetic3}
M. Cveti{\v{c}} and D. Youm, Nucl. Phys. {\bf B 472} (1996) 249.
\item
\label{cvetic4}
M. Cveti{\v{c}} and D. Youm, Nucl. Phys. {\bf B 476} (1996) 118.
\item
\label{tseytlin}
M. Cveti{\v{c}} and A.A. Tseytlin, Phys. Rev. {\bf D53 } (1996) 5619.
\item
\label{hull}
M. Cveti{\v{c}} and C.M. Hull, Nucl. Phys. {\bf B 480} (1996) 296.
\item
\label{sen2}
A. Sen, Int. J. Mod. Phys {\bf A9 } (1994) 3702.
\item
\label{penrose}
R. Penrose,
Riv. Nuovo Cimento {\bf 1} (1969) 252.
\item
\label{christodoulou}
D. Christodoulou,
Phys. Rev. Lett. {\bf 25} (1970) 1596.
\item
\label{gibbons}
G.W. Gibbons and C.M. Hull, Phys. Lett. {\bf B 109} (1982) 190.
\item
\label{auria}
L. Andrianopoli, R. D'Auria and S. Ferrara, hep-th 9612105.
\item
\label{duff}
M.J. Duff, J.T. Liu and J. Rahmfeld,
 Nucl. Phys. {\bf B 459} (1996) 125.
\item
\label{townsend}
C.M. Hull and P.K. Townsend, Nucl. Phys. {\bf B 438} (1995) 109.
\item
\label{scherk}
E. Cremmer, J. Scherk and J.H. Schwarz, Phys. Lett. {\bf B 84} (1979) 83.
\item
\label{cremmer}
E. Cremmer, Proceedings of the Gravity Workshop, Cambridge, June (1980) page
 267,
and Proceedings of the ICTP Spring School on Supergravity, Trieste, 
April (1981) page 313.
\item
\label{zumino}
B. Zumino, J. Math. Phys. {\bf 3} (1962) 1055.
\\
S. Ferrara, C. A. Savoy and B. Zumino,  
Phys. Lett. {\bf 100B} (1981) 393.
\item
\label{larsen}
F. Larsen and F. Wilczek,
Phys. Lett. {\bf B375} (1996) 37.
\end{enumerate}

\end{document}